\documentclass[twocolumn,aps,prd,showpacs,amsmath,amssymb,10pt,superscriptaddress]{revtex4-1}
\usepackage[dvips]{graphicx}
\usepackage{color}
\newcommand{\vect}[1]{\boldsymbol{#1}}

\begin{document}

\title{Thermal conductivity decomposition in two-dimensional materials:\\  Application to graphene}
\author{Zheyong Fan}
\email{brucenju@gmail.com}
\affiliation{COMP Centre of Excellence, Department of Applied Physics, Aalto University, Helsinki, Finland}
\author{Luiz Felipe C. Pereira}
\affiliation{Departamento de F\'{\i}sica, Universidade Federal do Rio Grande do Norte, Natal, RN, 59078-900, Brazil}
\author{Petri Hirvonen}
\affiliation{COMP Centre of Excellence, Department of Applied Physics, Aalto University, Helsinki, Finland}
\author{Mikko M. Ervasti}
\affiliation{COMP Centre of Excellence, Department of Applied Physics, Aalto University, Helsinki, Finland}
\author{Ken R. Elder}
\affiliation{Department of Physics, Oakland University, Rochester, Michigan 48309, USA}
\author{Davide Donadio }
\affiliation{Department of Chemistry, University of California at Davis, One Shields Avenue, Davis, California 95616, USA}
\author{Tapio Ala-Nissila}
\affiliation{COMP Centre of Excellence, Department of Applied Physics, Aalto University, Helsinki, Finland}
\affiliation{Department of Physics, Brown University, Box 1843, Providence, Rhode Island 02912, USA}
\affiliation{Department of Mathematical Sciences and Department of Physics, Loughborough University, Loughborough, Leicestershire LE11 3TU, UK}
\author{Ari Harju}
\affiliation{COMP Centre of Excellence, Department of Applied Physics, Aalto University, Helsinki, Finland}

\date{\today}

\begin{abstract}
Two-dimensional materials have unusual phonon spectra due to the presence of flexural (out-of-plane) modes. Although molecular dynamics simulations have been extensively used to study heat transport in such materials, conventional formalisms treat the phonon dynamics isotropically. Here, we decompose the microscopic heat current in atomistic simulations into in-plane and out-of-plane components, corresponding to in-plane and out-of-plane phonon dynamics, respectively. This decomposition allows for direct computation of the corresponding thermal conductivity components in two-dimensional materials. We apply this decomposition to study heat transport in suspended graphene, using both equilibrium and non-equilibrium molecular dynamics simulations. We show that the flexural component is responsible for about two thirds of the total thermal conductivity in unstrained graphene, and the acoustic flexural component is responsible for the logarithmic divergence of the conductivity when a sufficiently large tensile strain is applied.
\end{abstract}

\maketitle

\section{Introduction}

The high lattice thermal conductivity \cite{balandin2008,ghosh2008} of two-dimensional (2D) graphene and other carbon nanostructures has stimulated intensive studies to understand phonon transport in them \cite{balandin2011,marconnet2013,cepellotti2016}. Apart from holding great prospects for thermal management applications in nanoelectronic devices, graphene also serves as a benchmark for investigating fundamental questions regarding thermal transport in low-dimensional systems. Anomalous thermal transport, such as logarithmic divergence of thermal conductivity with respect to system size, has been long predicted for 2D lattice models \cite{lepri2003,yang2006,wang2012} and it has been debated whether or not this divergence can occur in graphene  \cite{xu2014,fugallo2014,barbarino2015,comeau2015apl,majee2016}. It has also been predicted that hydrodynamic phonon transport can occur in graphene in a much wider temperature range than in 3D materials \cite{lee2015,cepellotti2015}.  Moreover, effects of external conditions such as strain on the thermal transport in graphene have also attracted much attention \cite{bonini2012,pereira2013prb,lindsay2014,zhu2015,kuang2016}.

Because of the small anharmonicity in graphene, lattice dynamics methods based on perturbative treatments \cite{bonini2012,lindsay2014,fugallo2014,kuang2016} have been successfully used to study thermal transport in graphene. On the other hand, molecular dynamics (MD) based methods, which are nonperturbative, are also a valuable tool, especially in cases where the lattice dynamics-based methods are difficult to apply. Both the equilibrium MD (EMD) method based on the Green-Kubo formalism \cite{green1954,kubo1957} and the non-equilibrium  MD (NEMD) method \cite{mp1997,jund1999} based on Fourier's law have been extensively used. However, when used in their traditional form, little insight can be gained regarding the underlying transport mechanisms. There have been intensive efforts in developing MD-based methods for studying spectrally decomposed properties \cite{ladd1986,mcgaughey2004,henry2008,turney2009,larkin2014,feng2014,saaskilahti2014,saaskilahti2015,zhou2015,comeau2015,lv2016}, but most of them are targeted for general materials. One exception is the method by Gill-Comeau and Lewis \cite{comeau2015}, where the total thermal conductivity is decomposed into a single-particle component and a collective one, the latter being crucial to materials in which the non-resistive normal (non-umklapp) scattering is important, which is the case for graphene \cite{fugallo2014,lee2015,cepellotti2015}. Another recent development proposes that heat conduction in 2D materials is due to relaxons, which are wave packets of phonons that arise in the context of the linearized Boltzmann equation \cite{cepellotti2016}.

Here, we introduce an extension to the EMD and NEMD methods which is particularly useful for 2D materials. Specifically, we decompose the microscopic heat current into in-plane and out-of-plane parts, which are connected directly to the dynamics of the in-plane and the out-of-plane (flexural) phonons. Simulations based on this decomposition allow us to elegantly explore the effective time and length scales of phonon transport in strained as well as in unstrained graphene. Our results suggest that the thermal conductivity in unstrained pristine graphene is finite but diverges logarithmically when a sufficiently large tensile strain is applied. Combining the in-out decomposition and spectral decomposition, we find that the acoustic flexural phonon branch is responsible for the divergence of the thermal conductivity in pristine graphene under uni-axial tensile strain.

This paper is organized as follows.  Section \ref{section:theory} presents the theoretical formalisms used in this work. We formulate the in-out decomposition of the equilibrium heat current in the EMD method in Sec. \ref{section:green-kubo} and the in-out decomposition of the non-equilibrium heat current in the NEMD method in Sec. \ref{section:nemd-method}. In Sec. \ref{section:spectral}, the spectral decomposition method of S\"{a}\"{a}skilahti \textit{et al.} \cite{saaskilahti2014,saaskilahti2015} is generalized to include the in-out decomposition. Some details of our MD simulations are then presented in Sec. \ref{section:md}. After presenting the EMD and the NEMD results in Sec. \ref{section:emd-results} and Sec. \ref{section:nemd-results} respectively, we give a comparison between them in Sec. \ref{section:emd-vs-nemd}. Section \ref{section:summary} summarizes the results.

\section{Theory\label{section:theory}}

\subsection{Green-Kubo method\label{section:green-kubo}}

In the Green-Kubo method \cite{green1954,kubo1957,lepri2003}, the running lattice thermal conductivity along the $x$ direction can be expressed as
\begin{equation}
\label{equation:green-kubo}
\kappa_{xx}(t) = \frac{1}{k_BT^2V} \int_0^{t} dt' C_{xx}(t').
\end{equation}
Here, $k_B$ is Boltzmann's constant, $V$ is the volume of the system, $T$ is the absolute temperature, $C_{xx}(t')$ is the heat current autocorrelation function (HCACF), and $t$ is the correlation time. The HCACF is
\begin{equation}
 C_{xx}(t)=\langle J_{x}(0)J_{x}(t)\rangle,
\end{equation}
where $J_{x}(0)$ and $J_{x}(t)$ are components of the total heat current of the system at two time instants separated by an interval $t$. The symbol $\langle \rangle$ represents an ensemble average, which in EMD simulations equals the time average.

The heat current at a given time depends on the positions and velocities of the particles in the system. For many-body potentials, the calculation of the microscopic heat current is a highly nontrivial task  \cite{schelling2002,chen2006,Mandadapu2009,howell2012}. Recently, a well-defined expression valid for a general classical many-body potential has been derived as \cite{fan2015}
\begin{equation}
\label{equation:J}
\vect{J} = \sum_i \sum_{j \neq i} \vect{r}_{ij}
\left(
\frac{\partial U_j}{\partial \vect{r}_{ji}} \cdot \vect{v}_i
\right),
\end{equation}
where $\vect{r}_{ij} \equiv \vect{r}_{j} - \vect{r}_{i}$ is the position difference from atom $i$ to atom $j$ and
\begin{equation}
U_j = \frac{1}{2} \sum_{k \neq j} U_{jk},
\end{equation}
$U_{jk}$ being the bond energy between atoms $j$ and $k$. The heat current formula above is equivalent to that derived by Hardy at the quantum level \cite{hardy1963} and can be reexpressed in a more symmetric form:
\begin{equation}
\label{equation:J_symmetric}
\vect{J} = -\frac{1}{2} \sum_i \sum_{j \neq i} \vect{r}_{ij}
\left(
\frac{\partial U_i}{\partial \vect{r}_{ij}} \cdot \vect{v}_j -
\frac{\partial U_j}{\partial \vect{r}_{ji}} \cdot \vect{v}_i
\right).
\end{equation}
For two-body potentials, it reduces to
\begin{equation}
\label{equation:J_pair}
\vect{J}^{\text{two-body}} = -\frac{1}{4} \sum_i \sum_{j \neq i} \vect{r}_{ij}
\left[
\vect{F}_{ij} \cdot \left(\vect{v}_i + \vect{v}_j \right)
\right],
\end{equation}
where $\vect{F}_{ij}$ is the force on particle $i$ due to particle $j$.
As demonstrated in Ref. \cite{fan2015}, applying Eq. (\ref{equation:J_pair}) to 2D materials described by many-body potentials significantly underestimates the thermal conductivity.

The dot product in Eq. (\ref{equation:J}) can be decomposed into three terms, which correspond to the dynamics in different directions. In a three-dimensional isotropic system, all the three components contribute  equally.  However, in 2D systems, the in-plane and out-of-plane components are expected to have distinct characteristics.  This motivates a decomposition of  the heat current into an in-plane (the $x$-$y$ plane) component and an out-of-plane one,
\begin{equation}
\vect{J}=\vect{J}^{\text{in}}+\vect{J}^{\text{out}},
\end{equation}
where
\begin{equation}
\label{equation:J_in}
\vect{J}^{\text{in}} = \sum_i \sum_{j \neq i} \vect{r}_{ij}
\left(
\frac{\partial U_j}{\partial x_{ji}}  v_{xi} +
\frac{\partial U_j}{\partial y_{ji}}  v_{yi}
\right),
\end{equation}
and
\begin{equation}
\label{equation:J_out}
\vect{J}^{\text{out}} = \sum_i \sum_{j \neq i} \vect{r}_{ij}
\left(
\frac{\partial U_j}{\partial z_{ji}}  v_{zi}
\right).
\end{equation}
These two terms correspond to the contribution of in-plane and out-of-plane (flexural) phonon branches, respectively. With the heat current decomposition, we can define the following components of the HCACF:
\begin{equation}
C_{xx}= C_{xx}^{\text{in}} + C_{xx}^{\text{out}} + C_{xx}^{\text{cross}},
\end{equation}
where
\begin{equation}
C_{xx}^{\text{in}} = \langle J^{\text{in}}_{x}(0) J^{\text{in}}_{x}(t) \rangle;
\end{equation}
\begin{equation}
C_{xx}^{\text{out}} =
\langle J^{\text{out}}_{x}(0) J^{\text{out}}_{x}(t) \rangle;
\end{equation}
and
\begin{equation}
C_{xx}^{\text{cross}} =
2\langle J^{\text{in}}_{x}(0) J^{\text{out}}_{x}(t) \rangle.
\end{equation}
According to the decomposition above, the running thermal conductivity can also be decomposed into three terms:
\begin{equation}
\kappa^{\text{in}}_{xx}(t) =
\frac{1}{k_BT^2V} \int_0^{t} dt' C^{\text{in}}_{xx}(t') ;
\end{equation}
\begin{equation}
\kappa^{\text{out}}_{xx}(t) =
\frac{1}{k_BT^2V} \int_0^{t} dt' C^{\text{out}}_{xx}(t') ;
\end{equation}
\begin{equation}
\kappa^{\text{cross}}_{xx}(t) =
\frac{1}{k_BT^2V} \int_0^{t} dt' C^{\text{cross}}_{xx}(t').
\end{equation}

\subsection{Non-equilibrium molecular dynamics method\label{section:nemd-method}}

In the NEMD method, the system is driven out of equilibrium and when steady state is achieved, one measures the heat current (flux) and the temperature gradient from which one calculates the thermal conductivity of a sample with finite length $L$ according to Fourier's law. There are various versions of the NEMD method. First, the system can either have fixed \cite{saaskilahti2015,park2013,Mortazavi2016a,Pereira2016} or periodic boundary conditions \cite{mp1997,jund1999,schelling2002,howell2012,xu2014} along the transport direction. Second, the non-equilibrium heat current can be generated by different methods, including the velocity rescaling method by Jund and Jullien \cite{jund1999}, the velocity-swapping method by M\"{u}ller-Plathe \cite{mp1997}, or the thermostat method \cite{lepri2003,saaskilahti2015,park2013,Mortazavi2016a,Pereira2016}. It has been found that the results do not sensitively depend on the methods chosen (see e.g. \cite{xu2014}). To this end, we choose the periodic setup and generate the non-equilibrium heat current by using the method of Jund and Jullien \cite{jund1999}.

\begin{figure}[htb]
\begin{center}
\includegraphics[width=\columnwidth]{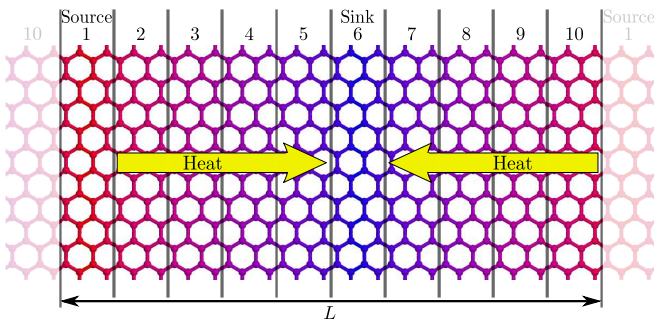}
\caption{A schematic illustration of the NEMD simulation setup. Here, a graphene sample of length $L$ is divided into $M=10$ blocks (separated by the vertical lines), labelled from 1 to $10$. The subsystem 1 acts as a heat source where energy flows in and the subsystem $M/2+1=6$ acts a heat sink where energy flows out. Periodic boundary conditions are applied in both planar directions of the graphene sheet. Therefore, heat flows from the source to the sink in two opposite directions, as indicated by the two arrows pointing towards the sink.}
\label{figure:nemd_setup}
\end{center}
\end{figure}

The system is divided along the transport direction into $M$ (an even number) blocks (labelled from 1 to $M$), with for instance block 1 chosen as a heat source and block $M/2+1$ as a heat sink, as schematically shown in Fig. \ref{figure:nemd_setup}.  The whole system is first equilibrated to a target temperature, and then the heat source/sink is heated/cooled with a given power $Q^{\text{ext}}=dE/dt$ for a sufficiently long time. After achieving steady state, one can start to record the block temperatures and heat flux.
The temperature of each block is calculated from the average kinetic energy of the atoms in that block via the equipartition theorem.
After obtaining the temperature gradient $|\nabla T|$ from the block temperatures, the conductivity of the (finite) system can be calculated according to Fourier's law as
\begin{equation}
\kappa(L) =
\frac{Q^{\text{ext}}/2}{S|\nabla T|},
\end{equation}
where $S$ is the cross-sectional area and the non-equilibrium heat current here should be taken as $Q^{\text{ext}}/2$ because the heat flows from the source to the sink in two opposite directions in the periodic boundary setup.

In the method of Jund and Jullien \cite{jund1999}, the non-equilibrium heat current $Q^{\text{ext}}$ can be externally controlled. Meanwhile, the non-equilibrium heat current can be expressed in terms of microscopic degrees of freedom. Following a procedure similar to that in Ref. \cite{fan2015}, the rate of energy increase of particle $i$ can be derived as
\begin{equation}
\frac{d E_i}{dt} = \sum_{j \neq i}
\left\langle
\left(\frac{\partial U_i}{\partial \vect{r}_{ij}} \cdot \vect{v}_j
-\frac{\partial U_j}{\partial \vect{r}_{ji}} \cdot \vect{v}_i\right)
\right\rangle.
\end{equation}
According to energy conservation, this rate should equal the sum of the rates $Q_{i\leftarrow j}$ of heat transfer from other particles
\begin{equation}
\frac{d E_i}{dt} = \sum_{j \neq i} Q_{i\leftarrow j}.
\end{equation}
Comparing the two equations above, we have
\begin{equation}
Q_{i\leftarrow j} = -Q_{i \rightarrow j} =
\left\langle
\left(\frac{\partial U_i}{\partial \vect{r}_{ij}} \cdot \vect{v}_j
-\frac{\partial U_j}{\partial \vect{r}_{ji}} \cdot \vect{v}_i\right)
\right\rangle.
\end{equation}
The total heat current from a block $A$ to another block $B$ is thus
\begin{align}
  Q_{A \rightarrow B}
  &=
  \sum_{i \in A} \sum_{j \in B} Q_{i \rightarrow j}
   \nonumber
  \\
  &=
-\sum_{i \in A} \sum_{j \in B}
\left\langle
\left(\frac{\partial U_i}{\partial \vect{r}_{ij}} \cdot \vect{v}_j
-\frac{\partial U_j}{\partial \vect{r}_{ji}} \cdot \vect{v}_i\right)
\right\rangle.
\label{equation:Q_AB}
\end{align}
This formula applies to general many-body potentials. For two-body potentials, it reduces to the following one:
\begin{equation}
  Q_{A \rightarrow B}^{\text{two-body}} =
- \frac{1}{2} \sum_{i \in A} \sum_{j \in B}
\left\langle \vect{F}_{ij} \cdot (\vect{v}_i + \vect{v}_j) \right\rangle.
\end{equation}

As in the case of the EMD simulations, we decompose the microscopic non-equilibrium heat current between two blocks $Q_{A \rightarrow B}$ into in-plane and out-of-plane components,
\begin{equation}
Q_{A \rightarrow B} = Q_{A \rightarrow B}^{\text{in}} + Q_{A \rightarrow B}^{\text{out}},
\end{equation}
where (using $r_{xij} \equiv x_{ij}$)
\begin{equation}
\label{equation:Q_in}
Q^{\text{in}}_{A \rightarrow B} = -\sum_{i \in A} \sum_{j \in B}
\left\langle
\sum_{\alpha=x,y}
\left(
\frac{\partial U_i}{\partial r_{\alpha ij}} v_{\alpha j}-
\frac{\partial U_j}{\partial r_{\alpha ji}} v_{\alpha i}
\right)
\right\rangle,
\end{equation}
and
\begin{equation}
\label{equation:Q_out}
Q^{\text{out}}_{A \rightarrow B} = -\sum_{i \in A} \sum_{j \in B}
\left\langle
\frac{\partial U_i}{\partial z_{ij}}  v_{zj}-
\frac{\partial U_j}{\partial z_{ji}}  v_{zi}
\right\rangle.
\end{equation}
Using the decomposed non-equilibrium heat current, we define the in-plane and out-of-plane thermal conductivities of a finite-length sample as
\begin{equation}
\kappa^{\text{in/out}}(L) =
\frac{Q^{\text{in/out}}}{S|\nabla T|}.
\end{equation}

We note that there is an important difference between the equilibrium heat current $\vect{J}$ defined in Eq. (\ref{equation:J}) and the non-equilibrium one $Q_{A \rightarrow B}$ defined in Eq. (\ref{equation:Q_AB}). The former fluctuates around zero in equilibrium and generally cannot be used in a non-equilibrium state, as demonstrated by Chen and Diaz \cite{chen2016}, while the latter equals the externally generated heat current $Q^{\text{ext}}/2$ in steady state, as shown in Appendix \ref{appendix:steady_state}. The non-equilibrium heat current expression we derived should be essentially equivalent to the formalism proposed by Chen and Diaz \cite{chen2016}, which could be used to find the spatial distribution of the heat flux at any time.

\subsection{Spectral decomposition\label{section:spectral}}

The non-equilibrium heat current can be further decomposed for different frequencies, as recently demonstrated by S\"a\"askilahti \textit{et al.} \cite{saaskilahti2014,saaskilahti2015}. Here, we extend their method to include the in-out decomposition. We first define the time-correlation functions $K^{\text{in/out}}_{A \rightarrow B}(t)$, which reduce to $Q^{\text{in/out}}_{A \rightarrow B}$ at $t=0$:
$K^{\text{in/out}}_{A \rightarrow B}(0)=Q^{\text{in/out}}_{A \rightarrow B}$. The out-of-plane part is defined as
\begin{equation}
\label{equation:K_out}
K^{\text{out}}_{A \rightarrow B}(t)
= \sum_{i \in A} \sum_{j \in B}
\left\langle
\frac{\partial U_i}{\partial z_{ij}}(0) v_{zj}(t)-
\frac{\partial U_j}{\partial z_{ji}}(0) v_{zi}(t)
\right\rangle,
\end{equation}
and the in-plane part is defined accordingly. These time-correlation functions are related to their Fourier transformed functions $\tilde{K}^{\text{in/out}}_{A \rightarrow B}(\omega)$ by
\begin{equation}
\label{equation:K}
K^{\text{in/out}}_{A \rightarrow B}(t) =
\int_{-\infty}^{\infty} \frac{d\omega}{2\pi} e^{-i\omega t}
\tilde{K}^{\text{in/out}}_{A \rightarrow B}(\omega),
\end{equation}
and
\begin{equation}
\tilde{K}^{\text{in/out}}_{A \rightarrow B}(\omega)=
\int_{-\infty}^{\infty} dt e^{i\omega t}
K^{\text{in/out}}_{A \rightarrow B}(t).
\end{equation}
Then, by setting $t=0$ in Eq.~(\ref{equation:K}) and noticing that $K^{\text{in/out}}_{A \rightarrow B}(-t)=K^{\text{in/out}}_{A \rightarrow B}(t)$, we arrive at the following spectral decomposition of the non-equilibrium heat current:
\begin{equation}
\label{equation:q_omega}
Q^{\text{in/out}}_{A \rightarrow B} =
\int_{0}^{\infty} \frac{d\omega}{2\pi}
\left[2\tilde{K}^{\text{in/out}}_{A \rightarrow B}(\omega)\right]
\equiv \int_{0}^{\infty} \frac{d\omega}{2\pi}
q^{\text{in/out}}_{A \rightarrow B}(\omega).
\end{equation}
After obtaining the spectral heat current $q^{\text{in/out}}_{A \rightarrow B}(\omega)$, one can calculate the spectral conductance per unit area ($\Delta T$ is the temperature difference between the source and the sink),
\begin{equation}
\label{equation:g_omega}
g^{\text{in/out}}_{A \rightarrow B}(\omega) =
\frac{q^{\text{in/out}}_{A \rightarrow B}(\omega)}{S |\Delta T|},
\end{equation}
and the spectral conductivity,
\begin{equation}
\label{equation:kappa_omega}
\kappa^{\text{in/out}}_{A \rightarrow B}(\omega) =
\frac{q^{\text{in/out}}_{A \rightarrow B}(\omega)}{S |\nabla T|}.
\end{equation}

\subsection{Details of the molecular dynamics simulations\label{section:md}}

We performed all the MD simulations using GPUMD (Graphics Processing Units Molecular Dynamics) \cite{fan2013,fan2015,fan2016}, an MD code which attains high performance on graphics processing units. To model the interactions between the carbon atoms, we use the Tersoff potential \cite{tersoff1989} with optimized parameters for graphene \cite{lindsay2010}.

In all the MD simulations, the velocity-Verlet integration method \cite{swope1982} with a time step of 1 fs is used for time-stepping. Energy conserves better than $10^{-5}$ in the microcanonical ensemble. Periodic boundary conditions are applied to both in-plane directions. All the simulations are performed at 300 K. The Berendsen barostat \cite{berendsen1984} is used to determine the equilibrium lattice constant. For pristine graphene, using the isothermal-isobaric ensemble, we found that the equilibrium lattice constant with zero stress at 300 K is slightly smaller than the value at zero temperature, which is a sign of negative thermal expansion due to the formation of thermal ripples \cite{michel2015,los2016}. After determining the room temperature lattice constant at zero stress in a square-shaped sample, we use it to simulate unstrained graphene in all other cases without controlling the stress any more. For strained graphene, we calculate the lattice constant in the strained direction according to the amount of applied strain. The definition of the thickness of 2D materials for reporting the effective three-dimensional thermal conductivity is arbitrary. In order to make close comparison with existing works, we use the conventional thickness of 0.335 nm.

\section{EMD results\label{section:emd-results}}

We first use the EMD method to compute thermal conductivity in pristine graphene. The simulation cell size is about 25 nm $\times$ 25 nm ($24~000$ atoms), which has been shown to be large enough to eliminate finite-size effects \cite{pereira2013prb,comeau2015,fan2015} in the Green-Kubo method.

\subsection{Thermal conductivity components in pristine graphene}

\begin{figure*}[htb]
\begin{center}
\includegraphics[width=2\columnwidth]{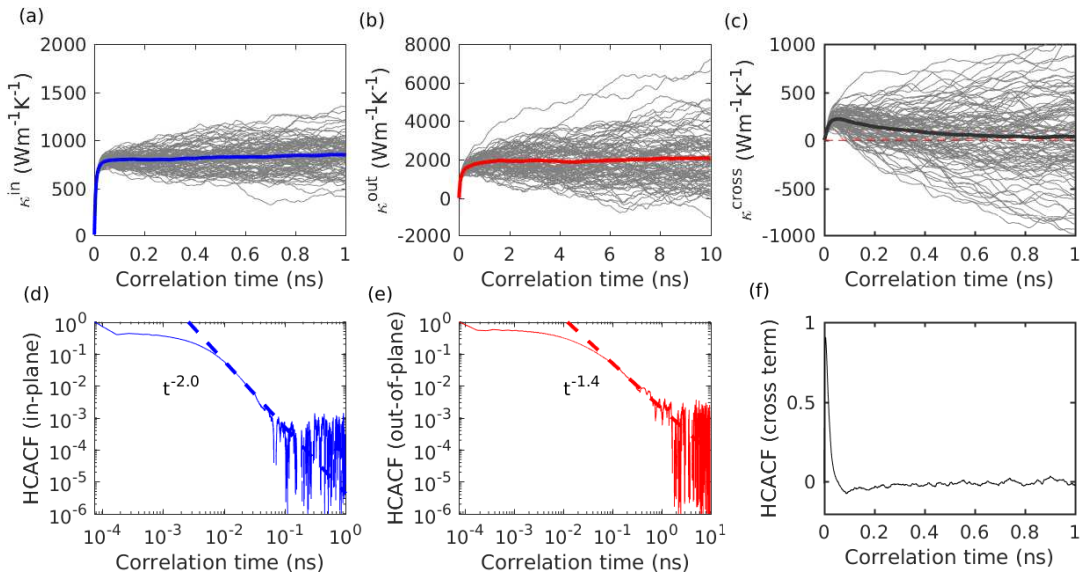}
\caption{ (a) The in-plane component, (b) the out-of-plane component, and (c) the cross term  of the running conductivity $\kappa(t)$ for pristine graphene at 300 K. The thin lines are results from 100 independent simulations and the thick lines represent the averages over the independent simulations. (d) The in-plane component, (e) the out-of-plane component, and (f) the cross term  of the normalized HCACF for pristine graphene at 300 K. The solid lines represent the raw data and the dashed lines are fits. See text for details. }
\label{figure:rtc_graphene}
\end{center}
\end{figure*}

In Ref. \cite{fan2015}, the running thermal conductivity $\kappa(t)$ of graphene  was computed at 300 K, but only up to a maximum correlation time of $t_{\text{max}}=$ 0.5 ns. As pointed out by Gill-Comeau and Lewis \cite{comeau2015}, this $t_{\text{max}}$ is not large enough to observe a complete saturation of the running conductivity. Below, we show that there are actually two distinct time scales governing the time-convergence of the running conductivity, and one of them exceeds 0.5 ns.

Figure \ref{figure:rtc_graphene} shows the calculated thermal conductivity components and the corresponding HCACFs for pristine graphene (averaged over the two in-plane directions) at room temperature. Here, we  consider a large maximum correlation time of $t_{\text{max}}=$ 10 ns. Since the fluctuations of the correlation function become larger with increasing correlation time, an extensive sampling in the phase space is required to obtain accurate results for large correlation times. The computational effort here is unprecedented: there are 100 independent simulations and each simulation lasts 51 ns (1 ns for equilibration and 50 ns for production), summing up to  $5.1~\mu s$.

Mode-coupling theory \cite{lepri2003} predicts a divergent $t^{-1}$ scaling of the HCACF for strictly 2D systems and a convergent $t^{-3/2}$ scaling for 3D systems. As shown in Fig. \ref{figure:rtc_graphene}(d) and (e), we find a best fit of $\sim t^{-2.0}$ for the in-plane component and $\sim t^{-1.4}$ for the out-of-plane component, which means that both components eventually saturate and $\kappa$ for pristine graphene is finite, in agreement with several recent theoretical studies using other approaches \cite{comeau2015apl,comeau2015,park2013,fugallo2014,barbarino2015,majee2016}, although it is found experimentally that $\kappa$ is still increasing up to 9 microns \cite{xu2014}.
Our results show clearly that the slow convergence of the thermal conductivity is due to the flexural phonons: the convergence of $\kappa^{\text{out}}(t)$ takes a few ns, while $\kappa^{\text{in}}(t)$ converges within a few hundred ps.

It is also important to note in Fig. \ref{figure:rtc_graphene}(a) and (b) that $\kappa^{\text{out}}(t)$ converges to a significantly larger value than $\kappa^{\text{in}}(t)$. Quantitatively, the in-plane and out-of-plane components converge to  $\kappa^{\text{in}}_0 \approx 850$ $\text{W}\text{m}^{-1}\text{K}^{-1}$ and
$\kappa^{\text{out}}_0 \approx 2~050$ $\text{W}\text{m}^{-1}\text{K}^{-1}$, respectively. The cross term  shows a peculiar behaviour, Fig. \ref{figure:rtc_graphene}(c), which is similar to a localization phenomenon, and is caused by the different time scales of the in-plane and out-of-plane phonons. Within a short correlation time, there is positive correlation between the two components and $\kappa^{\text{cross}}(t)$ reaches a peak value of about 200 $\text{W}\text{m}^{-1}\text{K}^{-1}$; at larger correlation time, the correlation between the two components is negative and $\kappa^{\text{cross}}(t)$ decays to zero. Asymptotically, $\kappa^{\text{cross}}_0$ can thus be taken as zero and we get a total thermal conductivity of
$\kappa^{\text{tot}}_0=2~900 \pm 100 $ $\text{W}\text{m}^{-1}\text{K}^{-1}$, where the error estimate is taken as the standard error of the independent runs.

\subsection{Strain effects}

\begin{figure}[hbt]
\begin{center}
\includegraphics[width=\columnwidth]{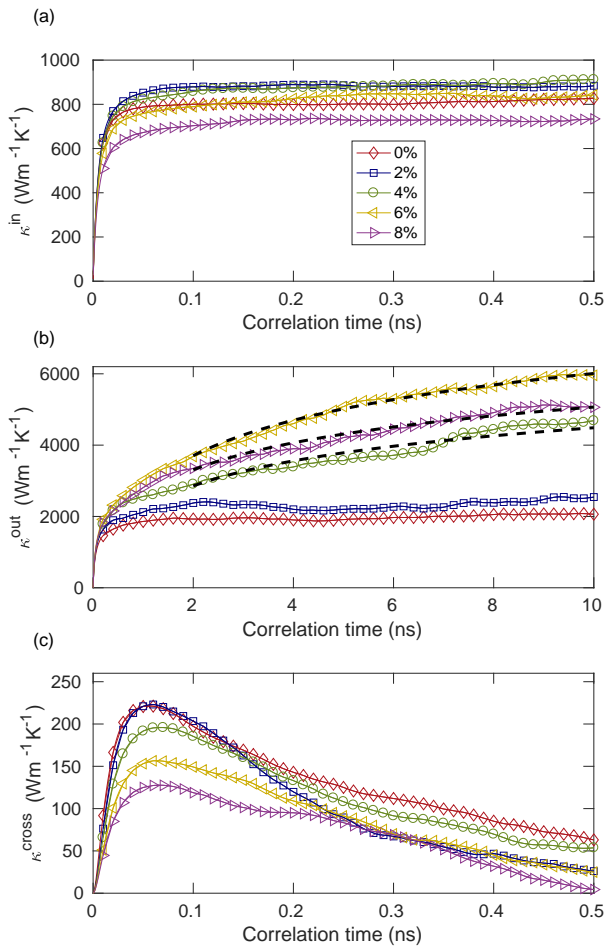}
\caption{(a) The in-plane component, (b) the out-of-plane component, and (c) the cross term  of the running thermal conductivity $\kappa(t)$ along the strained direction in pristine graphene under uni-axial tensile strain. The dashed lines represent fits to the data using a $\log(t)$ function.}
\label{figure:strain}
\end{center}
\end{figure}

Strain is usually unavoidable in real materials or it can be intentionally engineered \cite{Mohiuddin2009}. Figure \ref{figure:strain} shows the calculated running conductivity components in pristine graphene under uni-axial tensile strain. The results shown are obtained by applying the strain along the armchair direction, but similar results are obtained by applying the strain along the zigzag direction. The amount of strain $\epsilon$ was varied from $0\%$ (unstrained case) to a very large value of $8\%$. In strained graphene, the thermal conductivity is anisotropic, and here we are interested in the strained direction. We find that both the in-plane and the out-of-plane components of the conductivity perpendicular to the strained direction are reduced, as has also been found in previous works \cite{pereira2013prb,comeau2015}. In the following, we focus on the transport along the strained direction.

A striking difference between the behaviour of the in-plane and out-of-plane components can be seen: the in-plane component shows a increasing-to-decreasing trend with increasing strain, while the out-of-plane component becomes logarithmically divergent with respect to $t$ when $\epsilon>2\%$. The cross term  in strained graphene still shows localization and the peak value of the running conductivity decreases with increasing strain  when $\epsilon>2\%$.
The effect of divergence in the out-of-plane component becomes most prominent when $\epsilon = 6\%$, where the running conductivity at 10 ns shows a three-fold enhancement compared to that in unstrained graphene. Converting the time-divergence to length-divergence \cite{lepri2003}, we can conclude that thermal conductivity of pristine graphene under tensile strain diverges logarithmically with respect to the system size.

The divergence of $\kappa$ in strained graphene was first predicted \cite{bonini2012} to occur for any amount of strain based on first-principles lattice dynamics calculations using the single-mode relaxation time approximation, but previous \cite{pereira2013prb,comeau2015} and our current MD simulations indicate that the divergence does not occur when $\epsilon \lesssim 2\%$. In agreement with Ref. \cite{pereira2013prb}, we see that the out-of-plane phonon modes are responsible for the divergence. Our results also agree qualitatively with predictions based on full iterative solution of the linearized Boltzmann-Peierls equation for small ($<1\%$) \cite{lindsay2014} and large \cite{kuang2016} values of strain. In turn, the observed divergence for large strain disagrees with the results reported in Ref. \cite{fugallo2014}, which predict small changes of $\kappa$ for 4\% (biaxial) strain.

Regarding the physical origin of the divergence, Rold\'an \textit{et al.} \cite{roldan2011} have shown that anharmonic effects in stiff 2D membranes such as graphene can be dramatically suppressed by applying a tensile strain. They have considered biaxial (isotropic) tensile strain and found that a strain less than 1\% is sufficient to suppress the anharmonic coupling between bending and stretching modes in graphene, as evidenced by the flattening of the normal-normal correlation function $q^2G(q)$ in the region of small $q$ (large phonon wavelength). When the anharmonic effects are suppressed, the flexural phonons experience reduced scattering, causing the divergence of $\kappa$. We have also confirmed that $\kappa$ in graphene under 1\% biaxial tensile strain is already divergent (data not shown), which means that biaxial strain is more effective than uniaxial strain in suppressing the anharmonic effects.

\section{NEMD results\label{section:nemd-results}}

To gain more insight, we complement the EMD results above with NEMD simulations. After testing the convergence of $\kappa$ with the width of the simulation cell, we fix the width to 10 nm and consider samples of the following lengths: 0.2, 0.4, 0.8, 1.6, 3.2, 6.4, and 12.8 $\mu$m. The number of atoms ranges from 76 800 to 2 457 600. In all the NEMD simulations, the total simulation time for a given sample length is 25 ns: we first equilibrate the whole system under the target temperature for 5 ns and then switch on the external heat current for 15 ns, after which we record the block temperatures and the non-equilibrium microscopic heat current for 5 ns.

We have checked that stationary non-equilibrium conditions with a steady heat flux have been achieved in all the NEMD simulations. The temperature gradients in the simulations are also sufficiently small such that the assumption of linear response is valid; see the Appendices for details.

\begin{figure}[hbt]
\begin{center}
\includegraphics[width=\columnwidth]{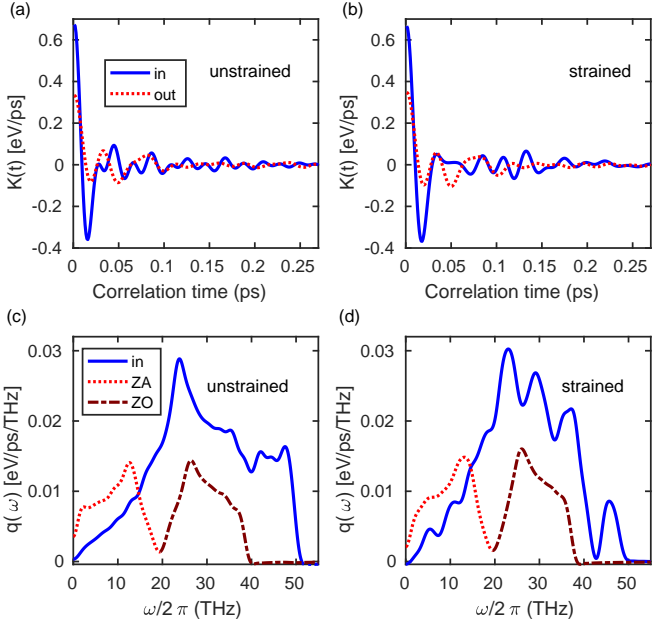}
\caption{(a-b) The correlation function $K^{\text{in/out}}_{A\rightarrow B}(t)$ [defined in Eq. (\ref{equation:K_out})] and (c-d) the spectral heat current $q^{\text{in/out}}_{A\rightarrow B}(\omega)$ [defined in Eq. (\ref{equation:q_omega})] in a short sample with zero or 6\% uni-axial tensile strain. The temperature is 300 K in both cases.}
\label{figure:spectral}
\end{center}
\end{figure}

Before presenting the NEMD results for the samples with different lengths, we first note that the non-equilibrium heat current components can be further spectrally decomposed \cite{saaskilahti2015}. The correlation function $K^{\text{in/out}}_{A\rightarrow B}(t)$ and the spectral heat current $q^{\text{in/out}}_{A\rightarrow B}(\omega)$ in a quasi-ballistic (20 nm long excluding the heat source and sink) sample, with or without strain, is shown in Fig.~\ref{figure:spectral}. The quasi-ballistic conductance $g^{\text{in/out}}_{A\rightarrow B}(\omega)$ defined by Eq. (\ref{equation:q_omega}), which is essentially the product of the phonon density of states and group velocity, is closely related to the phonon band structures. Noticeably, there is a high-frequency cutoff at $\sim 40$ THz and a band node at $\sim 20$ THz for the flexural modes in unstrained graphene, agreeing with the dispersion relations obtained by using the same empirical potential  \cite{pereira2013prb}. The band node for the flexural phonons also exists in strained graphene, allowing for distinguishing the flexural acoustic (ZA) from the flexural optical (ZO) modes in both unstrained and strained graphene.

\begin{figure}[hbt]
\begin{center}
\includegraphics[width=\columnwidth]{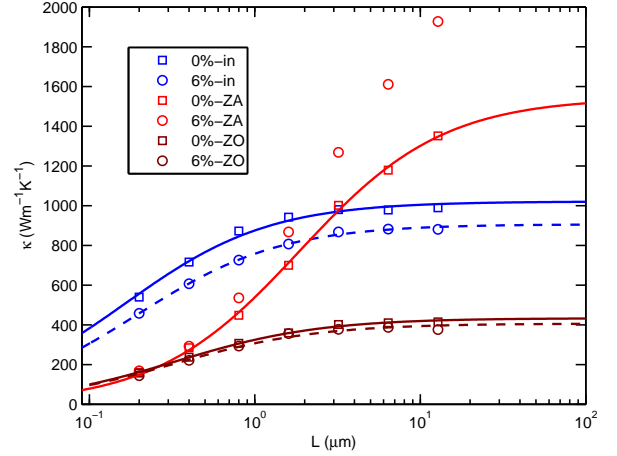}
\caption{Thermal conductivity components in unstrained
and strained (6\% uni-axial tensile strain) graphene as a function of the sample length $L$. The lines (solid lines for unstrained graphene and dashed lines for strained graphene) are fits to the corresponding markers according to Eq. (\ref{equation:ballistic-to-diffusive}) with $L_{\text{eff}}=L$, taking $\kappa_0^{i}$ and $\lambda^{i}$ as fitting parameters. The data for the ZA branch in strained graphene do not show saturation and the fit is omitted.}
\label{figure:nemd}
\end{center}
\end{figure}

With the help of the spectral decomposition, we can further calculate the length-dependent conductivity components $\kappa^{i}(L)~(i=\text{in,~ZA,~ZO})$. Here, we consider pristine graphene with zero and $6\%$ uni-axial tensile strain.  As shown in Fig.~\ref{figure:nemd}, all the components but $\kappa^{\text{ZA}}(L)$ in strained graphene show a trend of convergence with increasing $L$, in agreement with the EMD results. While the EMD results show that the flexural modes are the origin of the logarithmic divergence of conductivity in strained graphene, the NEMD results here provide evidence that the ZA modes are the ultimate source of the divergence.

For all the convergent components, the length dependence can be well described by the ballistic-to-diffusive crossover formula \cite{datta1995,wang2006}:
\begin{equation}
\kappa^{i}(L) \approx
\frac{\kappa_0^{i}}{1 + \lambda^{i}/L_{\text{eff}}}~ (i=\text{in,~ZA,~ZO}),
\label{equation:ballistic-to-diffusive}
\end{equation}
where $\lambda^{i}$ are the effective mean free paths (MFPs) of the different components and $\kappa^{i}_0$ are the corresponding thermal conductivities in the limit of infinite length. For the fixed boundary setup \cite{lepri2003,park2013,saaskilahti2015,Mortazavi2016a,Pereira2016}, where the source and sink are at the two ends of the sample and separated by $L$, it is clear that $L_{\text{eff}}=L$. For the periodic boundary setup \cite{mp1997,jund1999,schelling2002,howell2012,xu2014} used in this work, where the source and sink are separated by $L/2$, one usually uses $L_{\text{eff}}=L/2$. The exact value of $L_{\text{eff}}$ only affects the fitted effective MFPs. The fitted values of $\kappa_0^{i}$ are not affected by the value of $L_{\text{eff}}$ and are determined to be $\kappa_0^{\text{in}}\approx1020$ Wm$^{-1}$K$^{-1}$, $\kappa_0^{\text{ZA}}\approx1550$ Wm$^{-1}$K$^{-1}$, and $\kappa_0^{\text{ZO}}\approx430$ Wm$^{-1}$K$^{-1}$ for unstrained graphene. Their sum, $\kappa_0^{\text{tot}}\approx 3~000$ Wm$^{-1}$K$^{-1}$ in the infinite size limit, is consistent with the total conductivity obtained by the EMD method above. Taking $L_{\text{eff}}$ as $L$, the corresponding fitted effective MFPs are $\lambda^{\text{in}} \approx 170$ nm, $\lambda^{\text{ZA}} \approx 1~900$ nm, and $\lambda^{\text{ZO}} \approx 330$ nm, which would have been halved if $L_{\text{eff}}$ were taken as $L/2$.

\begin{figure}[hbt]
\begin{center}
\includegraphics[width=\columnwidth]{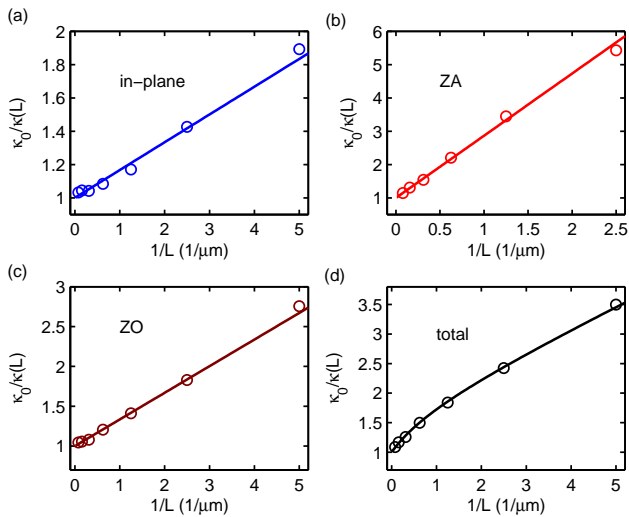}
\caption{Normalized inverse conductivity  $\kappa_0^{\text{in}}/\kappa^{\text{in}}(L)$ (a), $\kappa_0^{\text{ZA}}/\kappa^{\text{ZA}}(L)$ (b), $\kappa_0^{\text{ZO}}/\kappa^{\text{ZO}}(L)$ (c), and $\kappa_0^{\text{tot}}/\kappa^{\text{tot}}(L)$ (d) as a function of inverse length $1/L$. The lines in the subplots represent Eq. (\ref{equation:ballistic-to-diffusive}).}
\label{figure:k_inverse}
\end{center}
\end{figure}

We stress that a single effective MFP is a crude representation of the transport length scales for the different phonons in a given component/branch. However,
Eq. (\ref{equation:ballistic-to-diffusive}) with multiple effective MFPs already gives a significantly improved description of the data compared to the commonly used single-MFP formula \cite{schelling2002},
\begin{equation}
\kappa^{\text{tot}}(L) \approx
\frac{\kappa_0^{\text{tot}}}{1 + \lambda^{\text{tot}}/L_{\text{eff}}},
\label{equation:single_MFP}
\end{equation}
where $\lambda^{\text{tot}}$ is the effective MFP of all the phonons.
This can be seen from Fig. \ref{figure:k_inverse}, where the normalized inverse conductivity $\kappa_0/\kappa(L)$ is plotted as a function of the inverse length for the individual components (a-c) as well as their sum (d). While a linear relation between $\kappa_0/\kappa(L)$ and $1/L$ is followed for the individual components, the total conductivity shows a strong nonlinear behavior, which deviates from Eq. (\ref{equation:single_MFP}) but can be well described by Eq. (\ref{equation:ballistic-to-diffusive}). We note that this nonlinear behavior only shows up in very long samples ($L \gtrsim$ 1 $\mu$m), as has also been observed in the work of Park \textit{et al.} \cite{park2013}.

\section{Comparing EMD and NEMD results\label{section:emd-vs-nemd}}

In this work, both the EMD and NEMD methods are used, and it is important to make a closer comparison between them. To this end, we translate \cite{lepri2003} the time dependence in the EMD results into a length dependence using appropriate effective phonon group velocities $v^{i}~(i=\text{in},\text{out})$,
\begin{equation}
L \approx v^{i} t,
\end{equation}
and compare the EMD and NEMD data directly.

\begin{figure}[htb]
\begin{center}
\includegraphics[width=\columnwidth]{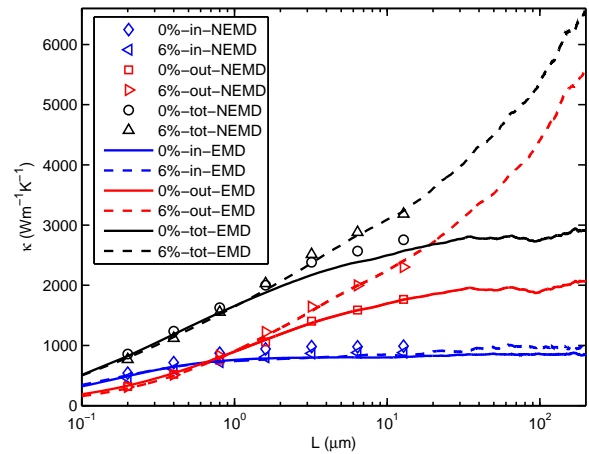}
\caption{Thermal conductivity components for graphene without strain (labelled by 0\% in the legend) and with 6\% uni-axial tensile strain obtained by NEMD and EMD simulations. The relation $L=v^i t$ ($i=$ in, out) with appropriate values of $v^i$ is used to convert the correlation time $t$ to an effective sample length $L$ in order to match the EMD data with the NEMD data.}
\label{figure:emd_vs_nemd}
\end{center}
\end{figure}

The comparison is shown in Fig. \ref{figure:emd_vs_nemd}. A good match between the EMD and NEMD data can be obtained by treating the group velocities as free parameters, which are fitted to be $v^{\text{in}} = 36$ km/s and $v^{\text{out}} = 21$ km/s for unstrained graphene, and $v^{\text{in}} = 28$ km/s and $v^{\text{out}} = 28$ km/s for graphene with 6\% uni-axial tensile strain. Similar to the effective MFPs, these effective group velocities are rough estimates. However, the general trend is clear: applying a tensile strain reduces $v^{\text{in}}$ and enhances $v^{\text{out}}$, which means that tensile strain softens the in-plane phonons but hardens the out-of-plane phonons. The fact that we need to treat the group velocities as fitting parameters may be justified in terms of the concept of relaxons proposed by Cepellotti and Marzari \cite{cepellotti2016}. The large relaxation times of the flexural modes observed in our MD results should be related to the relaxation times of relaxons whose velocities are not the same as the phonon velocities.

The interplay of these two effects can result in diverse strain effects \cite{zhu2015,kuang2016} on the thermal conductivity. When $L \lesssim 2$ $\mu$m, the softening of the in-plane phonons dominates and $\kappa$ can be decreased (slightly) by applying tensile strain. When $L \gtrsim 2$ $\mu$m, the hardening of the out-of-plane phonons dominates, which enhances $\kappa^{\text{tot}}$ and eventually makes it divergent with increasing sample length. The EMD data show that the divergence of $\kappa^{\text{tot}}$ is at least valid up to 200 $\mu$m. At this length scale, $\kappa^{\text{tot}}$ in strained graphene exceeds 6 000 Wm$^{-1}$K$^{-1}$, which is more than two times as large as that in unstrained graphene. For finite size patches, Fig. \ref{figure:emd_vs_nemd} shows that the difference between strained and unstrained systems is small, agreeing with the picture outlined by Fugallo \textit{et al.} \cite{fugallo2014} obtained by solving the Boltzmann transport equation of phonons. However, in contrast with our findings, they did not predict a divergent conductivity in the limit of infinite size. We note that Kuang \textit{et al.} \cite{kuang2016} predicted, also by solving the Boltzmann transport equation of phonons, that the conductivity of graphene diverges with increasing system size, even at high temperatures.

\section{Discussion}

Before concluding, we make some further remarks on our results.

\subsection{Comparison with previous works}

As pointed out in Ref. \cite{fan2015}, the heat current formula as
implemented in the popular MD package LAMMPS \cite{lammps_green_kubo} is incorrect for the Tersoff potential and results in significant underestimation of $\kappa$ using the Green-Kubo method. As most previous works have used LAMMPS, it is not straightforward to compare our results with them. One exception is Ref. \cite{comeau2015}, where LAMMPS was used to perform the MD simulations, with the correct Hardy formula \cite{hardy1963} (in the harmonic approximation) instead of the heat current formula as implemented in LAMMPS. The heat current formula by Hardy is identical to our exact heat current formula, as has been proven in Ref. \cite{fan2015}. Therefore, our method includes both the single-particle and collective components as defined in Ref. \cite{comeau2015}.
Qualitatively, the out-of-plane component in our formalism roughly (but not exactly) corresponds to the collective term in Refs. \cite{comeau2015} and \cite{fugallo2014}. In view of this, we expect that our results should be consistent to those in Ref. \cite{comeau2015}. A comparison between Fig. 2 of Ref. \cite{comeau2015} and Fig. 4 of Ref. \cite{fan2015} shows that the agreement in the calculated running thermal conductivity up to a few hundred pecoseconds is excellent. However, we point out that in the fitting of the out-of-plane component (or the collective component as defined in Ref. \cite{comeau2015}) of the HCACF using a power-law function $t^{-p}$ ($p$ is positive), the parameter $p$ depends sensitively on which part of the data are fitted. The fitting was done in the region of $t=0.1\sim 10$ ns in the current work, but was done in the region $t<0.6$ ns in Ref. \cite{comeau2015}. Using a region with smaller $t$ can result in an underestimate of $p$ and an overestimate of the extrapolated $\kappa$. Indeed, Ref. \cite{comeau2015} reported an extrapolated $\kappa$
of $3~998$ Wm$^{-1}$K$^{-1}$, which is about 30\% larger than our value ($2~900 \pm 100$ Wm$^{-1}$K$^{-1}$) obtained by directly reaching the region with converged $\kappa$.

We also have checked that our NEMD results are in excellent agreement with those in Xu \textit{et al.} \cite{xu2014}. Our NEMD results are also consistent with those in Park \textit{et al.} \cite{park2013}. There are some quantitative differences, though, which should be attributed to the different setups used in the NEMD simulations.

\subsection{Influence of quantum effects}

In our MD calculations, quantum effects were not taken in account. There is so far no reliable quantum correction to classical MD simulations available \cite{turney2009b}. Apart from giving a larger heat capacity, classical calculations also give shorter phonon lifetimes as compared to quantum calculations \cite{turney2009b}. According to lattice dynamics calculations \cite{singh2011}, these two opposite effects give rise to an overall 10\% underestimation of the thermal conductivity of graphene at room temperature. Usually, quantum corrections as applied in MD simulations only account for the quantum specific heat of the phonons, but not the quantum effects in the dynamics. This is also the case for some recently proposed mode-by-mode quantum correction methods in both EMD \cite{comeau2015,lv2016} and NEMD \cite{saaskilahti2016} simulations. Applying quantum corrections in this way only make the results deviate more from lattice dynamics calculations.

We do not consider quantum corrections in this study because our major goal is to propose the in-out decomposition method and give a direct comparison between EMD and NEMD results. Applying quantum corrections would mostly affect the in-plane part, which has relatively high phonon frequencies. Most importantly, the results for the ZA branch, which has relatively low phonon frequencies, would not be affected much and our conclusions regarding the length convergence/divergence will be still valid.

\section{Summary and Conclusions \label{section:summary}}

In summary, we have extended the formalisms of both EMD and NEMD simulations for thermal conductivity calculations by introducing a decomposition of the equilibrium and non-equilibrium heat currents, which allows for accessing the in-plane ($\kappa^{\text{in}}$) and out-of-plane ($\kappa^{\text{out}}$) components of the thermal conductivity $\kappa$ for 2D materials. We also demonstrated using the in-out decomposition in combination with spectral decomposition.

We have applied these methods to study heat transport in suspended pristine graphene. For unstrained pristine graphene, $\kappa$ was found to be upper-bounded and dominated by $\kappa^{\text{out}}$, which is about 2/3 of the total thermal conductivity. The scaling of thermal conductivity with respect to the sample length $L$ in pristine graphene can be well described by a simple ballistic-to-diffusive formula as expressed by Eq. (\ref{equation:ballistic-to-diffusive}).  When a uni-axial tensile strain exceeding 2\% is applied, the hardening of the ZA phonons results in a $\log(L)$ divergence of $\kappa$ with respect to the sample length $L$ in pristine graphene. Our results also show that the EMD and NEMD methods give consistent results for 2D materials and are largely complementary to each other.

The methods can also be applied to study heat transport in other 2D systems. Only homogeneous systems have been considered in this work, and it would be interesting to extend the formalisms to study interface heat transport in inhomogeneous systems. Computer implementation of the methods presented here will be made available in the near future through the GPUMD code \cite{fan2016}.

\begin{acknowledgments}
We thank helpful discussion with Kimmo S\"{a}\"{a}skilahti.
This research has been supported by the Academy of Finland through its Centres of Excellence Program (Project No. 251748).
We acknowledge the computational resources provided by Aalto Science-IT project and Finland's IT Center for Science (CSC).
ZF acknowledges the support of the National Natural Science Foundation of China (Grant No. 11404033).
LFCP acknowledges financial support from the Brazilian government agency CAPES for project ``Physical properties of nanostructured materials'' (Grant No. 3195/2014) via its Science Without Borders program.
PH acknowledges financial support from the Foundation for Aalto University Science and Technology.
KRE acknowledges financial support from the
National Science Foundation under Grant No. DMR-1506634.
\end{acknowledgments}

\appendix

\section{Steady state\label{appendix:steady_state}}

We have checked that a steady state has been fully achieved in all the NEMD simulations. This can be confirmed by two means.

\begin{figure}[htb]
\begin{center}
\includegraphics[width=\columnwidth]{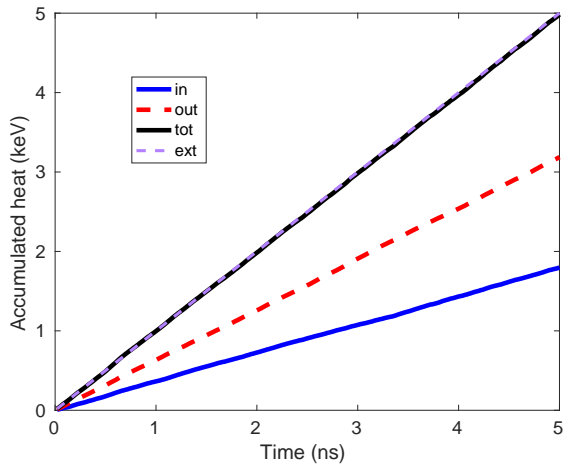}
\caption{Accumulated heat across a section as a function of time in the second longest sample (6.4 $\mu$m) in the NEMD simulation. The total heat (labelled as ``tot''), which is the sum of the in-plane component  (labelled as ``in'') and the out-of-plane component (labelled as ``out'') equals the energy externally supplied by the source (labelled as ``ext'').}
\label{figure:heat}
\end{center}
\end{figure}

On one hand, when steady state has been reached, the non-equilibrium heat current across an imaginary section should be equal to the power generated externally by the source and sink. This has been observed in all the samples. Figure \ref{figure:heat} shows an example in the second longest sample (without strain). Here, the out-of-plane component of the non-equilibrium heat current is larger. In shorter samples, the in-plane component can be larger. These features are reflected in the calculated thermal conductivity components. We note that S\"{a}\"{a}skilahti \textit{et al.} \cite{saaskilahti2015} and Gill-Comeau and Lewis \cite{comeau2015} have also demonstrated that applying the harmonic approximation in the calculation of the microscopic heat current barely affects the results. This is due to the small anharmonicity of the graphene lattice but cannot be guaranteed for other cases \cite{saaskilahti2014}.

\begin{figure}[htb]
\begin{center}
\includegraphics[width=\columnwidth]{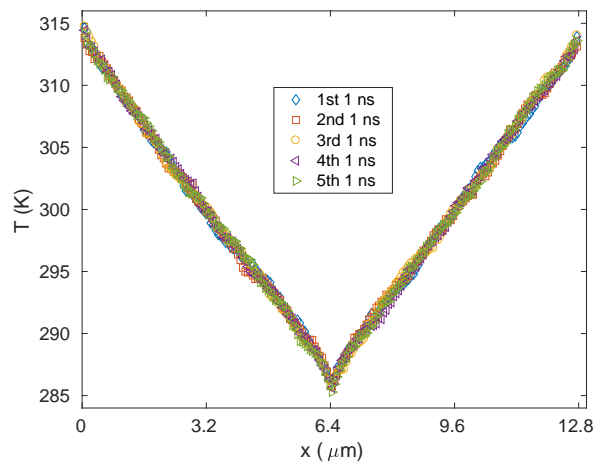}
\caption{Temperature profiles in the longest sample (12.8 $\mu$m) averaged
over different time intervals in the data-collecting stage. The data-collecting stage lasts 5 ns and each time interval lasts 1 ns.}
\label{figure:temp_time}
\end{center}
\end{figure}

On the other hand, when steady state has been reached, the temperature profile should not vary with time any more. This has also been observed in all the samples and Fig. \ref{figure:temp_time} shows an example in the longest sample (without strain): the temperature profiles averaged over the five one-nanosecond-long time intervals in the last 5 ns of the simulation do not show noticeable deviations from each other.

\section{Linear response}

After obtaining steady temperature profiles, we determine the temperature gradients by a linear fit, excluding the nonlinear regions around the source and sink. We stress that all the simulations are well within the linear-response regime of thermal transport, justifying the use of Fourier's law. Quantitatively, the temperature gradients we obtained range from about 0.06 K/nm in the shortest system to about 0.004 K/nm in the longest system, well below the upper bound of $\sim$ 1 K/nm, up to which linear response has been shown to be valid \cite{xu2014} for short systems. We note that in very long samples, the temperature gradient should be very small; otherwise, the temperature close to the heat source/sink would deviate significantly from the target temperature, resulting in non-linear temperature profiles.

\end{document}